\documentclass[12pt]{article}

\usepackage{graphics}
\usepackage{geometry}

\begin{document}

%

\title{Ultra High Energy Cosmic Rays and Air Shower Simulations:
a top-bottom view\date{} \footnote{
Talk presented by Claudio Corian\`{o}
at XV IFAE, Lecce, Italy, 23-26 April 2003}
\footnote{OUTP--03--16P}}

\author{$^{(1)}$A. Cafarella, $^{(1)}$C. Corian\`{o} and
$^{(2)}$A.E. Faraggi\\ \\
$^{(1)}$Dipartimento di Fisica,Universita' di Lecce,\\
INFN Sezione di Lecce,
Via Arnesano 73100, Lecce, Italy\\\\
$^{(2)}$Theoretical Physics Department,
University of Oxford, \\Oxford, OX1 3NP, United Kingdom}

\maketitle
\abstract{
Stable Superstring Relics (SSR) provide some of the candidates for
the possible origin of the Ultra High Energy Cosmic Rays (UHECR).
After a brief overview of the motivations for introducing such relics,
we address the question whether statistical fluctuations in the formation
of the air showers generated by the primary spectrum of protons can be
separated from a possible signal of new physics hidden in the first impact
with the atmosphere. Our results are generated by using minimal modifications
in the cross section of the primaries, and using available simulation codes
used by the experimental collaborations. The results indicate that
substantial increases in the cross section of the first impact, possibly due
to new interactions, are unlikely to be detected in geometrical and/or
variations of multiplicities in the cascade.
}

\section{Introduction}
Superstring theories represent the most advanced theories
to probe the unification of gravity and gauge interactions.
Over the past decade our understanding of the theory
has improved considerably with the realization
that the disparate string theories in ten dimensions,
together with eleven dimensional supergravity, are limits
of a more fundamental theory, dubbed M--theory. The question
remains however how to connect these advances to experimental data.
In this respect we may regard the ten dimensional string limits,
as well as the eleven dimensional limit, as effective limits of the
more fundamental theory. We should therefore anticipate that
non of these effective theories fully characterizes the true
nonperturbative vacuum, but can merely probe some of its
properties. Thus, for example, the heterotic limit is
the one that reveals the grand unification structures
that underly the Standard Model data, whereas the type I limit
may, perhaps, be more suited to study the dilaton stabilization issue.
In this context it may also be that the
confrontation of the fundamental theory with experimental data will be
achieved via its effective limits and its abstract formulation will
only be needed for conceptual consistency.

In this context one of the remarkable properties of the observed
Standard Model spectrum is the embedding of its matter multiplets
in spinorial representations of $SO(10)$. It is therefore sensible
to seek an effective string limit that preserves this embedding.
The only limit that realizes the $SO(10)$ embedding is the
heterotic limit as it is the only limit that gives rise to
spinorial representations in the perturbative spectrum. Indeed,
while a highly non--trivial task, phenomenological three
generation string models that preserve the $SO(10)$ embedding
have been constructed \cite{review}.

\section{Wilson-line Breaking Mechanism}
The prevalent method in string theory to break the non--Abelian
GUT symmetry is by utilizing Wilson line breaking. In turn, breaking of
the non--Abelian gauge symmetries in string vacua compactified
on non--simply connected manifolds results in massless
states that do not fall into representations of the original
unbroken GUT gauge group. The spectra thus contain states
that carry fractional charge with respect to the Abelian
Cartan generators of the original GUT gauge group \cite{fcs}.
Such states may carry fractional electric charge, or in models
in which the GUT symmetry is broken to $SU(3)_C\times SU(2)_L\times
U(1)_Y\times U(1)_{Z^\prime}$, the non--GUT states may carry the
standard charges under the Standard Model gauge group,
but a ``fractional'' non--GUT charge with respect to
to the additional $U(1)_{Z^\prime}$. This phenomena
is of primary importance for superstring phenomenology.
The main consequence is that it generically results in
super--massive states that are meta--stable. In the case
of fractionally charged states this is obvious. The
states are protected from decaying by electric charge
conservation. In the case of fractional $U(1)_{Z^\prime}$ charge
the meta--stability of the exotic state depends on the
charges of the Higgs spectrum that breaks the $U(1)_{Z^\prime}$
symmetry \cite{ccf}. Such states are therefore endemic in string
GUT models that utilize the Wilson--line symmetry breaking mechanism.

The typical mass scale of the exotic states will exceed the
energy range accessible to future collider experiments by several
order of magnitude. This general expectation follows from the fact that
the exotic states appear in vector--like representation
and therefore we in general expect that unsuppressed
mass terms are generated. In some cases the exotics mass
scale arises from a confinement scale of a hidden sector
non--Abelian gauge group. The question is therefore
whether the physics of the exotic states can be probed
experimentally, as it is unlikely to be accessible to collider
experiments.

In this respect one of the most intriguing experimental observations
of the past several decades is the observation of Ultra--High
Energy Cosmic Rays (UHECR) in excess of the so--called GZK
cutoff. Cosmic rays in excess of this bound are not expected
experimentally due to their scattering on the microwave
background, and consequent constraints on their mean free path.
Primaries with energy in excess of the GZK bound
could not have reached the earth from cosmological distances, whereas
there are no local astrophysical sources that can accelerate them
to the required energies. The meta--stable super--heavy
string relics suggest an appealing explanation for the observed
events in the form of so--called top--down models.
If the relics are sufficiently abundant in our local neighborhood,
and provided that their mass scale is of the right order,
then they can account for the observed events by their rare decay
into quarks and leptons. Thus, while the observation of
primaries with energies in excess of the GZK bound is still
an hotly debated experimental issue, this is clearly one of the
exciting possible experimental probes into the physics far
beyond that which is accessible to collider experiments.
The vital issue is therefore how to connect between the
data which is collected by the UHECR experiments and the
theoretical expectation of the heavy--relics from the string models
\cite{ccf}.
The string models predict specific states and quantum charges under
the four dimensional gauge group that can be modeled further
by using an effective field theory parameterization. Examples
of such states in concrete string models are provided
by the so called free fermionic heterotic string models \cite{review}.
\section{Can we detect new physics at Auger?}
There are various issues that can be addressed, both at theoretical and at
experimental level, on this point, one of them being an eventual confirmation
of the real existence of events above the cutoff.
However, even if these measurement will confirm their existence, remain yet to
be seen whether any additional new physics can be inferred just from an
analysis of the air shower.
A possibility might be supersymmetry \cite{coriano-faraggi}
or any new underlying interaction,
given the large energy available in the first impact.
We recall that the spectrum of the decaying X-particle
(whatever its origin may be),
prior to the atmospheric impact of the
UHECR is of secondary relevance, since the impact is always due to a single
proton.

What is relevant, instead, is the dynamics
in the evolution of the air shower and it is in the first collision that most
of the new channels
may become available.
One important point to keep into consideration is
that the new physical signal carried by the primaries in these collisions
is strongly ``diluted'' by their interaction with the atmosphere and that large
statistical fluctuations are immediately generated both by the randomness of
the first impact, the variability
in the zenith angle of the impact, and the -extremely large- phase space
available at those energies in terms of fragmentation channels.
\begin{figure}[tbh]
{\par\centering
\resizebox*{11cm}{!}{\rotatebox{-90}{\includegraphics{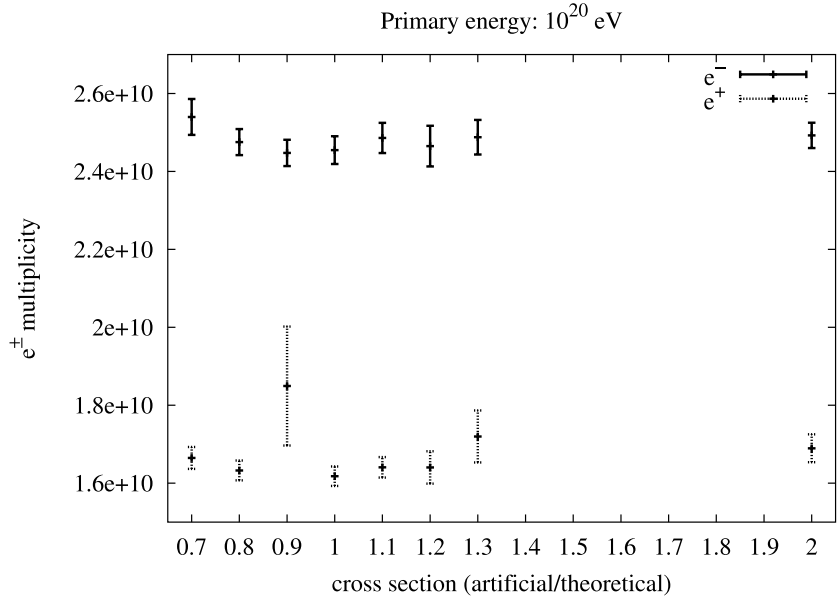}}} \par}
\caption{Multiplicities of $e^\pm$ for various values of the
correction factors modifying
the cross section of the first primary impact }
\end{figure}

\section{A simple test}
The simplest way to test whether a new interaction at the first proton-proton
impact can have any effect on the shower is to modify the cross section
at the first atmospheric impact using CORSIKA \cite{corsika} in combination
with
some
of the current hadronization models which are supposed to work at and around
the GZK cutoff. There are obvious limitations in this approach, since none of
the
existing codes incorporates any new physics beyond the standard model,
but this is possibly one of the simplest ways to proceed.
Therefore we take our results with some caution.

\begin{figure}[tbh]
{\par\centering
\resizebox*{11cm}{!}{\rotatebox{-90}{\includegraphics{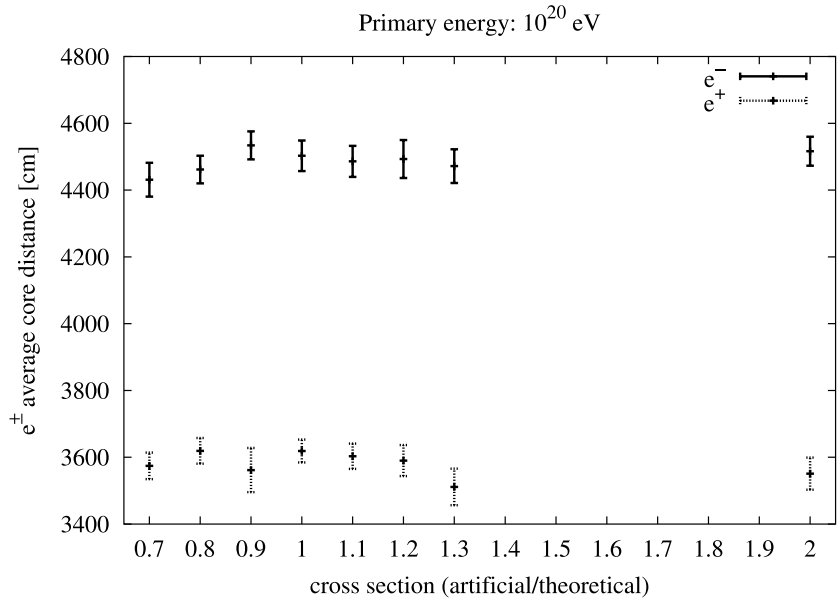}}} \par}
\caption{Averaged geometric (radial) distributions of $e^\pm$ measured with
respect
to the center of the detector (here located at 0 cm) for a zero zenith
impact of the primary proton versus the correction factors}
\end{figure}

We show in two figures results on the multiplicities,
obtained at zero zenith angle, of some selected particles
(electrons and positrons, in our case, but similar results hold
for all the dominant components of the final shower) obtained from a large
scale simulation of air showers at and around the
GZK cutoff. We have used the simulation code CORSIKA \cite{corsika} for this
purpose.

In Fig. 1 we show a plot obtained simulating an artificial first proton
impact in which we have modified the first interaction cross section by a
nominal factor ranging from 0.7 to 2. We plot on the y-axis the corresponding
fluctuations in the multiplicities both for electrons and positrons.
Statistical fluctuations \footnote{we keep the height of the first proton
impact with the
atmosphere arbitrary for each selected correction factor (x-axis)} have been
estimated
using bins of 60 runs. The so-developed showers have been thinned using the
Hillas
algorithm \cite{hillas}, as usually done in order to make the results of these
simulations manageable,
given the size of the showers at those energies. As one can immediately see,
the artificial
corrections on the cross section are compatible with ordinary fluctuations of
the air-shower.
The result is a negative one: a modified first impact, at least for such
correction factors
0.7-2 in the cross section of the first impact,
is unlikely to modify the multiplicities in any appreciable way.

A second test is illustrated in Fig. 2. Here we plot the same correction
factors on the
x-axis as in Fig. 1 but we show on the y-axis (for the same particles) the
average point
of impact on the detector and its corresponding statistical fluctuations.

As we increase the correction
factor statistical fluctuations in the formation of the air shower
seem to be compatible with the modifications induced by the ``new physics''
of the first impact and no special new effect is observed.

\section{Summary}
Fluctuations of these type, generated by a minimal
modification of the existing codes
only at the first impact may look simplistic, and can possibly
be equivalent to ordinary simulations with a simple rescale
of the atmospheric height at which the first collision occurs, since the
remaining
interactions are, in our approach, unmodified.
The effects we have been looking for, therefore, appear subleading compared to
other standard fluctuations which take place in the formation of the cascade.
On the other hand, drastic changes in the
structure of the air shower should possibly depend mostly on the physics
of the first impact and only in a less relevant way on the modifications
affecting the
cascade that follows up.
We have chosen to work at an energy of $10^{20}$ GeV but we do not
observe any substantial modifications of our results at lower energies
($10^{19}$ GeV), except for the multiplicities which are down by a factor
of 10. Our brief analysis, though simple, has the purpose to
illustrate one of the many issues which we believe
should be analyzed with great care in the near future:
the physics of the first
impact and substantial modifications to the existing codes
in order to see whether any new physics can be extracted from
these measurements from the multiplicities of the shower and its geometrical
shape. We have pointed out that it might be very difficult to disentangle any
new physics from
the large fluctuations of the air showers and set a warning over enthusiastic
claims such as detecting supersymmetry or other ``new'' interaction from these
types of studies. However, we do hope for the best, and clearly
improving our ability to extract viable information from cosmic
ray experiments continues to pose a vital challenge \cite{ACF}.

\section{Acknowledgments}

We thank D. Heck for discussions,
and D. Martello for discussions and for technical help.
Simulations have been performed using the INFN-LECCE.
computer farm. The work of A.F. is partly supported by PPARC.
The work of A.C. and C.C. is partly supported by INFN.

\end{document}